# Giant Hall effect in a highly conductive frustrated magnet $GdCu_2$

Kosuke Karube[1], Yoshichika Ōnuki[1], Taro Nakajima[1,2], Hsiao-Yi Chen[1,3], Hiroaki Ishizuka[4], Motoi Kimata[3], Takashi Ohhara[5], Koji Munakata[6], Takuya Nomoto[7], Ryotaro Arita[1,8], Taka-hisa Arima[1,9], Yoshinori Tokura[1,10,11] and Yasujiro Taguchi[1]

1. *RIKEN Center for Emergent Matter Science (CEMS), Wako 351-0198, Japan*
2. *The Institute for Solid State Physics, University of Tokyo, Kashiwa 277-8581, Japan*
3. *Institute for Materials Research, Tohoku University, Sendai 980-8577, Japan*
4. *Department of Physics, Tokyo Institute of Technology, Meguro-ku 152-8551, Japan*
5. *J-PARC Center, Japan Atomic Energy Agency, Tokai 319-1195, Japan*
6. *Comprehensive Research Organization for Science and Society, Tokai 319-1106, Japan*
7. *Department of Physics, Tokyo Metropolitan University, Hachioji 192-0397, Japan*
8. *Department of Physics, University of Tokyo, Bunkyo-ku 113-0033, Japan*
9. *Department of Advanced Materials Science, University of Tokyo, Kashiwa 277-8561, Japan*
10. *Department of Applied Physics, University of Tokyo, Bunkyo-ku 113-8656, Japan*
11. *Tokyo College, University of Tokyo, Bunkyo-ku 113-8656, Japan.*

**Abstract:**

**The Hall effect is one of the most fundamental but elusive phenomena in condensed matter physics due to the rich variety of underlying mechanisms. Here we report an exceptionally large Hall effect in a frustrated magnet $GdCu_2$ with high conductivity.**

**The Hall conductivity at the base temperature is as high as the order of $10^4$-$10^5$ $\Omega^{-1}cm^{-1}$ and shows abrupt sign changes under magnetic fields. Remarkably, the giant Hall effect is rapidly suppressed as the longitudinal conductivity is lowered upon increasing temperature or introducing tiny amount of quenched disorder. Our systematic transport measurements together with neutron scattering measurements, *ab initio* band calculations and spin model calculations indicate that the unusual Hall effect can be understood in terms of spin-splitting induced emergence/disappearance of Fermi pockets as well as skew scattering from spin-chiral cluster fluctuations in a field-polarized state. The present study demonstrates complex interplay among magnetization, spin-dependent electronic structure, and spin fluctuations in producing the giant Hall effect in highly conductive frustrated magnets with a distorted triangular lattice.**

## Introduction

The Hall effect in magnetic materials has attracted growing attention due to its diverse quantum physical mechanisms [1, 2]. Traditionally, the microscopic origins of the anomalous Hall effect (AHE) have been classified into two categories: intrinsic and extrinsic ones. The intrinsic mechanism of the AHE is attributed to the Berry curvature associated with the electron bands as affected by magnetic structures. Recently, large intrinsic AHEs due to non-collinear or non-coplanar spin textures have been extensively studied in antiferromagnetic metals [3−6]. In addition, large topological Hall effects (THE) arising from magnetic skyrmions, nanometric topological spin textures, have been found in Gd-based intermetallic compounds dominated by magnetic frustration [7−9].

The extrinsic AHE, on the other hand, has long been well known to originate from impurity scatterings, where highly conducting electrons are scattered asymmetrically from impurities under the influence of spin-orbit coupling [10, 11]. Recently, skew scattering from spin clusters was theoretically proposed as a new mechanism of the AHE [12, 13]. In this spin-cluster scattering, scalar spin chirality (SSC) derived from non-coplanar spin textures induces large Hall conductivities, and hence large Hall angles. This new mechanism of skew scattering explains the giant unconventional Hall effects in the frustrated kagome magnet $KV_3Sb_5$ [14] and the chiral magnet MnGe [15]. The recent progress in these unconventional Hall effects in the high-conductivity regime has stimulated scientific interest in the search for giant Hall effects in magnetic materials with both high conductivity and non-trivial spin textures.

Here, we report an exceptionally large Hall conductivity and its novel magnetic field response in a highly conductive frustrated magnet $GdCu_2$, which belongs to the family of $RCu_2$ ($R$: rare-earth elements) with the orthorhombic structure (space group: *Imma*) as shown in Fig. 1(a, b). This structure can be viewed as a distorted $AlB_2$-type hexagonal structure [16], and Gd atoms form distorted triangular lattice layers perpendicular to the *b*-axis. $GdCu_2$ has been known to exhibit antiferromagnetic ordering below $T_N$ = 40 K [17]. In 2000, using neutron scattering, Rotter *et al*. reported a non-collinear magnetic structure described by the modulation vector $\boldsymbol{Q}_0$ = (2/3, 1, 0), which corresponds to an in-plane cycloidal propagation along the *a*-axis with a pitch angle of 120° and an anti-parallel coupling along the *b*-axis, as depicted in Fig. 1b [18]. More recently, incommensurate magnetic order with a slightly modified *a*-axis propagation $\boldsymbol{Q}_0$ = (0.678, 1, 0) has been reported [19]. The magnetization along the *a*-, *b*- and *c*-axes shows the similar field dependence and exhibits two metamagnetic transitions accompanied by magnetostrictions around 6 T and 8 T [20−22]. Despite the observation of such a non-trivial magnetic structure and metamagnetic field response, there have been no experimental reports of Hall effects.

In this study, magnetic and electric transport properties of $GdCu_{2-x}Au_x$ ($0 \leq x \leq 0.05$) were investigated systematically using high-quality single crystals (see Supplementary Fig. S1 for details of sample characterizations). In addition, we conducted neutron scattering measurements and *ab initio* band calculations to clarify the magnetic and electronic structures under magnetic fields. In addition, we calculated spin chirality fluctuations in the field-induced ferromagnetic state using spin wave approximation. Using those comprehensive data, we discuss the plausible origin of the giant Hall effect in terms of band structure change and cluster skew scattering.

## Results

### Magnetic and transport properties

First, we show basic magnetic and transport properties of $GdCu_2$ in Fig. 1(c-f). In these measurements, the magnetic field ($H$) is applied parallel to the $b$-axis of the crystal (defined as $z$-axis) and the electric current ($J$) is applied along the $a$-axis (defined as $x$-axis), and thus the Hall voltage is measured along the $c$-axis (defined as $y$-axis). Temperature dependence of magnetic susceptibility $\chi$ shows a sharp cusp at $T_N$ = 40 K (Fig. 1c), indicating the 120° order of Gd moments. The Curie-Weiss fitting yields the Weiss temperature $\theta_p$ = 4.3 K and the effective magnetic moment $\mu_{eff}$ = 8.97 $\mu_B$/Gd, respectively. The positive Weiss temperature may imply the presence of ferromagnetic interaction in the system although the actual magnetic order is antiferromagnetic. As shown in Fig. 1d, the longitudinal conductivity $\sigma_{xx}$ shows an abrupt upturn below $T_N$ and reaches as high as 1.6 x $10^6$ $\Omega^{-1}cm^{-1}$ at 2 K (see Supplementary Fig. S3 for the temperature dependence of the longitudinal resistivity $\rho_{xx}$). Magnetic field dependence of the magnetization $M$ at 2 K exhibits two metamagnetic transitions around 6 T and 8 T and saturates around 10 T (Fig. 1e), in good agreement with the previous report [21]. Here, we call the magnetic phases as I (0-6 T), II (6-8 T), III (8-10 T), and induced ferromagnetic (FM) phase (> 10 T). The corresponding Hall conductivity $\sigma_{xy}$ at 2 K as a function of the field is shown in Fig. 1f. The Hall conductivity exhibits a large positive peak ($\sigma_{xy} \sim 4 \times 10^4\ \Omega^{-1}\mathrm{cm}^{-1}$) in phase I, followed by a sudden decrease in phase II, and then sharply drops in phase III, exhibiting large negative values ($\sigma_{xy} \sim -3 \times 10^4\ \Omega^{-1}\mathrm{cm}^{-1}$). As the magnetic field is further increased, the Hall conductivity rapidly increases and shows a positive peak ($\sigma_{xy} \sim 4 \times 10^4\ \Omega^{-1}\mathrm{cm}^{-1}$) again in the FM region.

**Thermal effect in high fields**

To clarify the behavior of the Hall conductivity up to higher magnetic fields and to investigate the influence of thermal fluctuations, we performed high-field resistivity measurements up to 24 T (see Methods for details). Figure 2a shows the field-swept $\sigma_{xy}$ at several elevated temperatures, and a contour plot of $\sigma_{xy}(T, H)$ on the phase diagram obtained from magnetization measurements (Supplementary Fig. S2) is displayed in Fig. 2b. At 2 K, huge positive maxima on the order of $10^5\ \Omega^{-1}\mathrm{cm}^{-1}$ are observed around 5 T (phase I) and 11 T (FM phase), and $\sigma_{xy}$ eventually decays with a broad tail as the magnetic field is further increased. Upon increasing temperature, both positive peaks in phase I and the FM phase are suppressed, but their temperature and field dependence are clearly distinct as described below. The large positive Hall conductivity in phase I decreases rapidly with increasing temperature but without changing the peak position, as indicated with the square symbols in Fig. 2(a, b), and shows small negative values above 8 K. On the other hand, the decrease in the large positive Hall conductivity in the FM region is more gradual and the peak position shifts to the high field side upon increasing temperature, as indicated with the triangle symbols in Fig. 2(a, b), and a broad positive peak is still observed above 10 K in high fields. These differences point to different origins of the large Hall conductivity in the low-field region and the FM region, namely, the normal Hall effect (NHE) of high-mobility carriers and the extrinsic AHE due to spin-chirality-induced skew scattering, respectively, as discussed later. Since the large $\sigma_{xy}$ in the FM region is most likely the extrinsic AHE, the maximum value of $\sigma_{xy}^{\max}$ at each temperature from 2 K to 12 K is plotted as a function of $\sigma_{xx}$ and compared to the conventional skew scattering of Fe films and the giant AHEs of $KV_3Sb_5$ [14] and MnGe [15] in Fig. 2c. The power law fitting yields a scaling relation of $\sigma_{xy} \propto \sigma_{xx}^{1.78}$.

**Influence of quenched disorder**

As mentioned above, the large Hall conductivity is observed only at low temperatures, where the longitudinal conductivity is very high on the order of $\sigma_{xx} \sim 10^6\ \Omega^{-1}\mathrm{cm}^{-1}$. Therefore, next we intentionally polluted the sample by doping a tiny amount of Au to investigate the dependence of the giant Hall effect upon quenched disorder. The longitudinal and Hall conductivities of $GdCu_{2-x}Au_x$ ($x = 0$, 0.02 and 0.05) are summarized in Fig. 3. As expected, $\sigma_{xx}$ decreases with increasing the Au concentration (Fig. 3a), while the magnetic properties are almost unchanged as confirmed by magnetization measurements (see Supplementary Fig. S3 for details). Figures 3b and 3c show the magnetic field dependence of $\sigma_{xx}$ and $\sigma_{xy}$, respectively, at 2 K for $x = 0$, 0.02 and 0.05 (see Supplementary Figs. S4 and S5 for the magnetic field dependence of $\rho_{xx}$ and $\rho_{yx}$ at all the measured temperatures). While the magnitude of $\sigma_{xx}$ decreases by a factor of 2-3 with increasing $x$, its field dependence is qualitatively unchanged. On the other hand, $\sigma_{xy}$ is more significantly suppressed with increasing $x$, and the characteristic field dependence with sign reversal completely disappears for $x = 0.05$. This Au-doping dependence of $\sigma_{xy}$ at 2 K bears similarity to the temperature dependence in the non-doped sample. Therefore, the giant Hall conductivity in $GdCu_2$ is intimately related to the magnitude of the longitudinal conductivity, as shown in the scaling plot of Fig. 2c.

**Magnetic structure under magnetic fields**

To search for a possible change in the magnetic structure of $GdCu_2$ under magnetic fields, we performed single-crystal neutron diffraction measurements (see Methods for details). Figure 4a shows the reciprocal space intensity mapping on the ($h$, 1, $l$) plane obtained at 2 K and 0 T. Magnetic reflections associated with $\boldsymbol{Q}_0 = (2/3, 1, 0)$ are observed clearly as indicated with green circles, which is consistent with the previous study [18]. The intensity line cuts along $h$ around (4/3, 1, 0) [= (2, 2, 0) − $\boldsymbol{Q}_0$] and (8/3, 1, 0) [= (2, 0, 0) + $\boldsymbol{Q}_0$] at various magnetic fields are plotted in Fig, 4b. These magnetic reflections are found to persist up to 6.5 T above the metamagnetic transition field around 5.5 T. The integrated intensities of these magnetic peaks are plotted as a function of magnetic field in Fig. 4c. As the magnetic field is increased, these intensities decrease above 5.5 T but a half of the intensity remains at 6.5 T. No other magnetic reflections arising from a different finite modulation $Q$ vector are observed at 6.5 T in the detectable range of the reciprocal lattice space (Supplementary Fig. S7). As schematically depicted in Fig. 4d, the Gd 4$f$ local moments lying in the $ac$-plane tilt to the $b$-axis by an angle $\theta$ as the magnetic field is increased. The above experimental results, in which the $Q$ vector is invariant up to 6.5 T, indicate that the metamagnetic transition from phase I to II is accompanied by the jump of $\theta$ while the in-plane component maintains the 120° structure and interlayer antiferromagnetic arrangement. Since the diffraction intensity is proportional to the square of the in-plane component of the magnetic moments, if the above scenario is correct, the intensity should be proportional to $\cos^2\theta$. In fact, as shown in Fig. 4c, the field dependence of the integrated intensity shows good agreement with the $\cos^2\theta$ calculated from the magnetization curve. Although 6.5 T is insufficient to reach phase III, $\theta$ is expected to further increase from phase II to III around 7.5 T, and finally the Gd moments are forcedly almost aligned to the *b*-axis above 10 T. As described in the theoretical section later, the conduction bands having large hybridization with Gd 5*d* orbitals split significantly with increasing magnetization due to strong coupling with the frustrated Gd 4*f* local moments. As a result, new Fermi surface of spin-polarized conduction electrons may emerge, accompanied by the abrupt increase in the tilt angle $\theta$ between the magnetic phases.

**Origin of the giant Hall effect**

Here, we discuss the plausible origin of the giant Hall conductivity from phase I to phase III. Since the upper limit of $\sigma_{xy}$ value based on the intrinsic mechanism of AHE is on the order of $10^3$ $\Omega^{-1}$cm$^{-1}$ [1], the intrinsic mechanism can hardly explain the experimental result of $\sigma_{xy}$ ~ $10^4$-$10^5$ $\Omega^{-1}$cm$^{-1}$. In addition, the giant Hall conductivity appears only when the longitudinal conductivity is very high ($\sigma_{xx}$ ~ $10^6$ $\Omega^{-1}$cm$^{-1}$). Therefore, the giant Hall conductivity can be attributed to either extrinsic AHE or normal Hall effect (NHE). For multiple carriers with high mobility, the NHE often exhibits a nontrivial magnetic field ($B$) dependence as described by the Drude model,

$$\sigma_{xy}^{(\mathrm{Drude})} = \sum_i \mu_i n_i q_i \cdot \frac{\mu_i B}{1 + (\mu_i B)^2}$$

where $\mu_i$, $n_i$ and $q_i$ are mobility, density and charge of each carrier species, respectively.

To investigate the NHE without the magnetic contribution, we performed similar transport measurements for a non-magnetic isostructural compound $YCu_2$ [23] without *f* moments and observed even higher longitudinal and Hall conductivities ($\sigma_{xx}$ ~ 3.6 x $10^6$ $\Omega^{-1}cm^{-1}$ and $\sigma_{xy} \sim 3 \times 10^5\ \Omega^{-1}cm^{-1}$) at 2 K (Supplementary Fig. S7). The Hall conductivity in $YCu_2$ shows an initial non-linear slope and a broad single peak as a function of the external field, and thus can be well reproduced with a two-carrier Drude model. Therefore, it is reasonable to assume that the large Hall conductivity in $GdCu_2$ is also governed by the NHE of high-mobility carriers arising from a similar band structure. In the case of $GdCu_2$, as discussed in the following theoretical section, the density as well as the sign of carriers vary significantly at the magnetic phase transitions, bringing additional features to the Hall conductivity.

**First-principles band structure calculations**

First-principles calculations utilizing density functional theory were performed to gain insights into the atypical behavior exhibited by the Hall conductivity concerning the external magnetic field and internal magnetization (see Methods for details). Since the nature of carriers determines the transport property, we plot the band structure near the Fermi level ($E_F$) and highlight the characteristics of the conduction electrons and holes which have strong hybridization with Gd 5*d* orbitals. Figure 5(a, b) shows band structures with $k_z = 0$ and $k_z = 0.5$ along a high-symmetry path within the *xy*-plane. By comparing the paramagnetic (PM) bands (blue line) and FM bands (red line), we notice that the magnetic effect induces a spin-splitting of 50-300 meV, which substantially changes the nature of the carriers on the Fermi level as the net magnetization is enhanced. The large spin splitting clearly demonstrates the strong coupling between the Gd 4*f* local spins and the conduction electrons in this material. Given that the observed changes in carrier density at the Fermi level represent a general response to magnetization changes, the subsequent discussion of the transition from PM to FM phases can be reasonably extended to more complex magnetic structure changes as observed in the experiment.

In the following, we categorize the bands by their spin-resolved electronic states at the $E_F$ in PM/FM states. There are four types, and we summarize them in Fig. 5(c). As in the standard approach, since the upward parabolic bands form an electron pocket when they cross the $E_F$, we call upward parabolic bands the "electron type" (Type-e); since the downward parabolic bands contribute as holes, we call downward parabolic bands the "hole type" (Type-h). Besides, for both electron and hole types, the bands that cross the $E_F$ in the PM state without spin splitting, but only for the majority spin band in the FM state with spin splitting, we call them type 1; for those that do not touch the $E_F$ in the PM state but come across it in the FM state, we call them type 2. With this categorization, we can discuss their contribution to the Hall conductivity as we increase the external magnetic field.

According to the Boltzmann transport equation, the electron ($q_i$ = $-e$) contributes negatively to the Hall conductivity while the hole ($q_i$ = $+e$) contributes positively. Therefore, for the band of Type-e1(h2), the spin-down band stops (starts) contributing as electrons (holes) to the transport property as the spin-splitting is enhanced by the external magnetic field. Therefore, the Hall conductivity increases as a response to the changing of carrier density for each carrier species. On the other hand, for the band of Type-e2(h1), the spin-up band starts (stops) contributing as electron(hole) to the transport as the spin-splitting is increased, resulting in a decrease in the Hall conductivity. In Fig. 5(a, b), we highlight some notable bands that belong to the four introduced types and attribute the magnetization dependence of Hall conductivity to the change of band structure near the $E_F$. Therefore, the sharp decrease in the Hall conductivity in phases II and III can be explained by the significant change in the Fermi surface, which is plausibly dominated by Type-e2 or Type-h1.

**Skew scattering by spin clusters**

In addition to the NHE dominated by the high-mobility carriers in the spin-polarized band structure in the phases I-III, below we discuss the possibility of the extrinsic AHE as the origin of the sharp increase in $\sigma_{xy}$ in the FM phase.

As mentioned above, the metamagnetic transition and the corresponding decrease in the Hall conductivity from phase I to II and from II to III are probably accompanied by changes in the Fermi surface. On the other hand, as the magnetization smoothly connects from phase III to the FM region, it is unlikely that such a significant change in the Fermi surface occurs around the saturation field. Therefore, we attribute this part to AHE. As shown in the scaling plot of AHE in Fig. 2c, the large magnitude of $\sigma_{xy} \sim 10^4$-$10^5$ $\Omega^{-1}cm^{-1}$ and its strong dependence on $\sigma_{xx}$ cannot be explained by the intrinsic mechanism. In the extrinsic mechanisms, i.e., conventional skew scattering and side jump, on the other hand, the anomalous Hall conductivity is proportional to the magnetization at a fixed temperature, and the anomalous Hall angle ratio ($\tan\theta_H = \sigma_{xy}/\sigma_{xx}$) is typically less than 1% [24]. Therefore, these conventional impurity scattering mechanisms can never explain the exceptionally large anomalous Hall angle ($\tan\theta_H \sim 12\%$ at 2 K) and unusual temperature-field dependence of the Hall conductivity in the field polarized state as shown in Fig. 2a, b. Thus, we can finally conclude that skew scattering by spin clusters with spin chirality fluctuations is the most plausible scenario. The neutron scattering measurement indicates that the *b*-axis magnetization process is caused by the increase in the canting angle of Gd moments from the *ac* plane while keeping the in-plane 120° modulation (Fig. 4d). Thus, spin chirality fluctuations near the phase boundary (local spin canting from a collinear state) may induce skew scattering. While the total spin chirality cancels out in a uniform regular triangle lattice, complete cancellation does not occur in a distorted triangular lattice as in $GdCu_2$, as discussed quantitatively in the following theoretical section.

**Model calculation of spin chirality**

To investigate how the scalar spin chirality of fluctuating spins occurs in the field-induced FM phase, we consider an effective triangular lattice model

$$H = -J_1 \sum_{\langle i\alpha, j\alpha\rangle_1} \vec{S}_{i\alpha}\cdot\vec{S}_{j\alpha} - J_2 \sum_{\langle i\alpha, j\beta\rangle_2} \vec{S}_{i\alpha}\cdot\vec{S}_{j\beta} - J_2' \sum_{\langle i\alpha, j\beta\rangle_2'} \vec{S}_{i\alpha}\cdot\vec{S}_{j\beta} + \sum_{\langle i\alpha, j\alpha\rangle_1} D_\alpha \hat{z}\cdot\vec{S}_{iA}\times\vec{S}_{jA} - h\sum_{i\alpha} S_{i\alpha}^z,$$

where $\vec{S}_{i\alpha} = (S_{i\alpha}^x, S_{i\alpha}^y, S_{i\alpha}^z)$ is the Gd spin on $\alpha(=A,B)$th sublattice in $i$th unit cell, and the summations are for 1, 2 and $2'$ bonds in Fig. 6a; the first three terms are antiferromagnetic Heisenberg interaction ($J_1$, $J_2$, $J_2' < 0$) between the Gd spins and the fourth term is the Dzyaloshinskii-Moriya (DM) interaction allowed by the symmetry. Unlike the uniform triangular lattice, the DM interaction is allowed in $GdCu_2$ by the lattice distortion, which breaks the inversion symmetry on bond 1. The DM interaction favors a spin configuration with opposite spin chirality for blue and red triangles. In addition, the inequivalence of blue and red triangles violates the cancellation of net spin chirality, which often prohibits a spin-chirality-related AHE in the uniform triangular lattice. Hence, the distortion in $GdCu_2$ can give rise to a non-zero spin chirality and chirality-related AHE.

We calculated the spin chirality of the above model in the FM phase using linear spin-wave theory. This theory is reliable in the high field region where the tilting of spins by the thermal fluctuation is small; as a threshold, we used $\langle S_{ix}^2 + S_{iy}^2\rangle \leq 1$, where $\langle S_{ix}^2 + S_{iy}^2\rangle$ is the thermal average of transverse spin components. The magnetic field dependence of spin chirality for the blue triangles is shown in Fig. 6b, which shows non-zero chirality at a finite temperature. Note that the spin chirality is induced by the fluctuating spins, not by a chiral magnetic order. In the presence of such chiral spin fluctuation, it induces a skew-scattering AHE proportional to the spin chirality, $\sigma_{xy} \propto \chi(T,h)$ [12]. The spin chirality increases as the magnetic field decreases, which is qualitatively consistent with the AHE curve in experiment [Fig. 2(a)]. In addition, in the low temperature, the spin chirality increases as the temperature increases (Supplementary Fig. S10), reproducing the temperature dependence in the high field region [Fig. 2(a, b)]. Hence, the temperature and magnetic field dependence of AHE in the FM phase is qualitatively reproduced by the skew-scattering AHE.

Triangular lattice magnets with 120° spin ordering have been less investigated than kagome lattice magnets in the studies of anomalous Hall effects because of the spin-chirality cancelation problem, although there are exceptions such as $PdCrO_2$ [25, 26]. Here we demonstrate that giant spin-cluster skew scattering occurs in triangular lattice magnets and that symmetry breaking due to the lattice distortion is important for the non-zero spin chirality. As shown in Fig. 2c, the maximum value of the anomalous Hall conductivity in $GdCu_2$ is even larger than those in the kagome magnet $KV_3Sb_5$ [14] and the chiral magnet MnGe [15] with similar mechanisms, exhibiting the largest value ever reported.

**Conclusions**

We performed comprehensive magnetic and transport measurements for the highly conductive frustrated magnet $GdCu_2$ and discovered the giant Hall conductivity on the order of $10^4$-$10^5$ $\Omega^{-1}cm^{-1}$, which shows an atypical response to the increase in net magnetization and decreases significantly as the longitudinal conductivity is lowered. This unconventional Hall effect is governed by the normal Hall effect of high mobility carriers where carrier densities vary significantly with increasing magnetization due to spin-splitting induced emergence/disappearance of the Fermi pockets. Furthermore, the characteristic positive peak of the Hall conductivity near the ferromagnetic phase boundary is presumably attributed to the anomalous Hall effect due to skew scattering from spin clusters accompanying scalar spin chirality. This study demonstrates the important roles of both band and spin structures in the Hall conductivity and contributes significantly to the understanding of the fundamental physics of the giant Hall effect in non-coplanar spin structure with high mobility carriers.

## Methods

### Sample preparations

Bulk single crystals of $GdCu_2$ and $GdCu_{2-x}Au_x$ ($x$ = 0.02 and 0.05) were prepared by the Bridgman method. First, polycrystalline samples were synthesized from pure Gd and Cu metals in the stoichiometric ratio by arc melting in an Ar atmosphere. The product was crushed into small pieces and loaded in an yttria crucible and sealed in an evacuated quartz tube. Then, the quartz tube was set in a vertical Bridgman furnace and slowly lowered in the region with a temperature gradient between the upper heater (910°C) and the lower heater (800°C) at a rate of 1.4 mm/h. The obtained single crystal ingot was cut into rectangular pieces after determining the crystal orientation with an X-ray Laue camera (RASCO-BL II, Rigaku). The $R$Cu$_2$-type orthorhombic crystal structure and phase purity were confirmed by powder X-ray diffraction with Cu K$\alpha$ radiation (RINT-TTR III, Rigaku) as shown in Supplementary Fig. S1.

### Magnetization measurements

Temperature-swept magnetization measurements were performed by a superconducting quantum interference device magnetometer (MPMS3, Quantum Design), while field-swept magnetization was measured up to 14 T using a physical properties measurement system (PPMS, Quantum Design) equipped with a vibrating-sample magnetometer (VSM) option. In all measurements, the external magnetic field was applied parallel to the *b*-axis. In the magnetization measurements displayed in Fig. 1e and Fig. 2b, the magnetic field was applied perpendicular to the sample plate so that the demagnetization factor ($N \sim 0.7$) is the same as that of the sample used for the transport measurements. Similarly, for the magnetization curve shown in Fig. 4c, the magnetic field was applied parallel to the sample plate so that the demagnetization factor ($N \sim 0.03$) is the same as that of the sample used for the neutron scattering measurements.

**Electric transport measurements**

Longitudinal resistivity ($\rho_{xx}$) and Hall resistivity ($\rho_{yx}$) were measured up to 14 T using a conventional four-terminal method in a PPMS (Quantum Design) equipped with an AC transport option. A plate-like single crystal with an area of 1.9 mm x 1.0 mm (*ac* plane) and a thickness of 0.3 mm (along the *b* axis) was used. In the AC transport measurements, the electric current (40 mA, 197 Hz) was applied along the *a*-axis and the external field was applied along the *b*-axis (perpendicular to the plate). To eliminate the influence of voltage probe misalignment, the raw data of field-swept longitudinal and Hall resistivity were processed as $\rho_{xx}(H) = [\rho_{xx}(+H) + \rho_{xx}(-H)]/2$ and $\rho_{yx}(H) = [\rho_{yx}(+H) - \rho_{yx}(-H)]/2$, respectively. The field-swept longitudinal and Hall conductivities were calculated with the formulae $\sigma_{xx}(H) = \rho_{xx}(H)/[\rho_{xx}^2(H) + \rho_{yx}^2(H)]$ and $\sigma_{xy}(H) = \rho_{yx}(H)/[\rho_{xx}^2(H) + \rho_{yx}^2(H)]$, respectively, where we assume $\rho_{yy} = \rho_{xx}$ and $\rho_{xy} = -\rho_{yx}$ Temperature-swept longitudinal conductivity was obtained from $\sigma_{xx}(T) \sim 1/\rho_{xx}(T)$.

**High-field transport measurements**

The high-field transport measurements of $GdCu_2$ up to 24 T were performed using a 25 T cryogen-free superconducting magnet (25T-CSM) in Institute for Materials Research (IMR), Tohoku University. The longitudinal and Hall resistivity values $\rho_{xx}$ and $\rho_{yx}$ were measured using lock-in amplifiers with AC electric current (35.4 mA). To eliminate the voltage probe misalignment, the resistivity values were averaged with the positive field sweep and the negative field sweep (measured after rotating the sample stage by 180°) as described above. These data were used in the calculation of the longitudinal and Hall conductivity $\sigma_{xx}$ and $\sigma_{xy}$ and plotted in Fig. 2.

**Neutron scattering measurements**

Neutron scattering measurements on $GdCu_2$ were performed using a time-of-flight (TOF) single crystal neutron diffractometer (SENJU, BL-18) at the Materials and Life Science Experimental Facility (MLF) of the Japan Proton Accelerator Research Complex (J-PARC) [27]. A plate-like single crystal with a large area of 9 mm x 5 mm (*bc*-plane) and a thickness of 0.4 mm (along the *a*-axis) was used. The magnetic field was applied parallel to the *b*-axis using a vertical-field superconducting magnet up to 6.5 T. By controlling the rocking angle ($\omega$) between the incident neutron beam (wavelength range: 0.4−4.4 Å) and the horizontal *c*-axis, the neutron diffraction data were accumulated for 3-4 hours for each rocking angle at $\omega$ = 12°, 30°, 47° and 102°. The data were processed and the intensity mapping on the reciprocal lattice space was obtained by the software STARGazer [28].

**First-principles calculations**

In the first-principles calculations for $GdCu_2$, the experimental lattice constants are set to $a$ = 4.320 Å, $b$ = 6.858 Å, and $c$ = 7.33 Å [16], with optimized atomic position with QUANTUM ESPRESSO code [29]. We use pseudopotentials with PBEsol exchange-correlation functional generated from the Standard Solid-State Pseudopotentials library (SSSP) [30, 31]. To take account of the strong correlation among electrons within the Gd atom, we add an empirical Hubbard *U* parameter of 7.0 eV on the 4*f*-orbitals. Using a converged energy cutoff of 70Ry with a 6 x 4 x 4 k-grid, we obtained a ferromagnetic (FM) ground state with 7.12 $\mu_B$/atom for Gd and -0.01 $\mu_B$/atom for Cu. To discuss the dependence on magnetization, we compute another band structure in addition to the FM band, where the exchange magnetic field is manually turned off in the Kohn-Sham equation to represent the paramagnetic (PM) states.

**Spin wave approximation in the FM phase**

We computed the scalar spin chirality using classical spin wave approximation. In the FM phase, at a low temperature, the Gd spins points along the *z* axis, i.e., $\vec{S}_{i\alpha} \sim (0,0,1)$. Hence, by assuming $S_{i\alpha}^{x,y} \ll 1$ and $S_{i\alpha}^{z} = \sqrt{1 - {S_{i\alpha}^{x}}^2 - {S_{i\alpha}^{y}}^2} \sim 1 - \frac{1}{2}({S_{i\alpha}^{x}}^2 + {S_{i\alpha}^{y}}^2)$, the Hamiltonian reads $H = \frac{1}{2}\sum_{\vec{k}} \vec{S}_{-\vec{k}} h_{\vec{k}} \vec{S}_{\vec{k}}$, where

$$\begin{aligned} h_{\vec{k}} = &(-2J_1 \cos(k_x) + 2J_1 + 2J_2 + 2J_2' + h_z)\sigma_0 \otimes \sigma_0 - i2D_z \sin(k_x)\, \sigma_3 \otimes \sigma_2 \\ &- 2(J_2 + J_2') \cos\left(\frac{\sqrt{3}k_y}{2}\right) \sigma_1 \otimes \sigma_0 + 2i(J_2 - J_2') \sin\left(\frac{\sqrt{3}k_y}{2}\right) \sigma_2 \otimes \sigma_0 \end{aligned}$$

with $\sigma_{1,2,3}$ being the Pauli matrices and $\sigma_0$ is the 2 x 2 unit matrix, $\vec{S}_{\vec{k}} = (S_{\vec{k}A}^{x}, S_{\vec{k}A}^{y}, S_{\vec{k}B}^{x}, S_{\vec{k}B}^{y})$, $S_{\vec{k}\alpha}^{a} = \frac{1}{\sqrt{N}}\sum_i S_{i\alpha}^{a} e^{-i\vec{k}\cdot\vec{r}_{i\alpha}}$, and $\vec{r}_{i\alpha}$ is the position of spin $\vec{S}_{i\alpha}$.

Similarly, the spin chirality for the blue triangles reads

$$\begin{aligned} \chi_1 = &\frac{1}{N}\sum_{\vec{r}} \hat{z}\cdot\left(\vec{S}_{B,\vec{r}} \times \vec{S}_{A,\vec{r}} + \vec{S}_{A,\vec{r}} \times \vec{S}_{A,\vec{r}+\vec{x}} + \vec{S}_{A,\vec{r}+\vec{x}} \times \vec{S}_{B,\vec{r}}\right) + \hat{z} \\ &\cdot\left(\vec{S}_{B,\vec{r}} \times \vec{S}_{A,\vec{r}+\vec{x}} + \vec{S}_{A,\vec{r}+\vec{x}} \times \vec{S}_{B,\vec{r}+\vec{x}} + \vec{S}_{B,\vec{r}+\vec{x}} \times \vec{S}_{B,\vec{r}}\right) \\ = &\frac{1}{N}\sum_{\vec{r}} i\sin(k_x)\left[S_{A-\vec{k}}^{x} S_{A\vec{k}}^{y} - S_{A-\vec{k}}^{y} S_{A\vec{k}}^{x} - S_{B-\vec{k}}^{x} S_{B\vec{k}}^{y} + S_{B-\vec{k}}^{y} S_{B\vec{k}}^{x}\right], \end{aligned}$$

and the magnetization is

$$M^z \sim 1 - \frac{1}{4N} \sum_{\vec{r},\alpha} \left[ \left( S^x_{\alpha\vec{r}} \right)^2 + \left( S^y_{\alpha\vec{r}} \right)^2 \right] = 1 - \frac{1}{4N} \sum_{\vec{k},\alpha} \left[ S^x_{\alpha-\vec{k}} S^x_{\alpha\vec{k}} + S^y_{\alpha-\vec{k}} S^y_{\alpha\vec{k}} \right].$$

Using these formulas, the thermal average of a physical observable $O = \sum_k \vec{S}_{-k} o_{\vec{k}} \vec{S}_k$, including chirality and magnetization, is given by

$$\langle O \rangle \sim \frac{1}{Z} 2^{N_{asym}/2} \int D\, \vec{S}_{\vec{k}} \frac{1}{N} \sum_k \vec{S}_{-k} o_{\vec{k}} \vec{S}_k \exp\left( -\frac{\beta}{2} \sum_k \vec{S}_{-\vec{k}} h_k \vec{S}_{\vec{k}} \right),$$

where

$$Z = tr\left( e^{-\beta H} \right) \sim 2^{N_{asym}} \int D\, \vec{S}_{\vec{k}} \exp\left( -\frac{\beta}{2} \sum_k \vec{S}_{-\vec{k}} h_k \vec{S}_{\vec{k}} \right)$$

is the partition function.

**Data availability**

All the data presented in the article and Supplementary Information are available from the corresponding authors upon reasonable request.

**Acknowledgments**

We are grateful to N. Nagaosa, M.-K. Lee and M. Mochizuki for fruitful discussions. We also thank A. Kikkawa for technical support for experiments. This work was supported by JSPS Grant-in-Aids for Scientific Research (Grant No. 23K26534, 23H04014, 23H04868, 23KK0052, 22H00109, 22H04933, 23K22447, 22H01176, 21H05470), JST

CREST (Grant No. JPMJCR20T1), and the RIKEN TRIP initiative (Many-body Electron Systems and Advanced General Intelligence for Science Program). This work was partly performed under the GIMRT program of the Institute for Materials Research, Tohoku University (Proposal No. 202311-HMKPA-0001). The neutron experiment at the Materials and Life Science Experimental Facility of the J-PARC was performed under a user program (Proposal No. 2022B0259).

**Author contributions**

K.K., Y.O., T.A., Y. Tokura and Y. Taguchi jointly conceived the project. K.K. and Y.O. synthesized bulk crystals. K.K. performed magnetization and transport measurements. High-field transport measurements were carried out by K.K. and M.K.. Neutron scattering measurements were performed by K.K., T.N., T.O. and K.M.. H.C., T.N. and R.A. conducted the band calculations. H.I. performed the spin wave approximation calculations. The results were discussed and interpreted by all the authors. K.K., H.C., H.I. and Y. Taguchi wrote the manuscript.

**Additional Information**

Correspondence and requests for materials should be addressed to K. Karube and Y. Taguchi.

**Competing financial interests**

The authors declare no competing financial interests.

## Figures

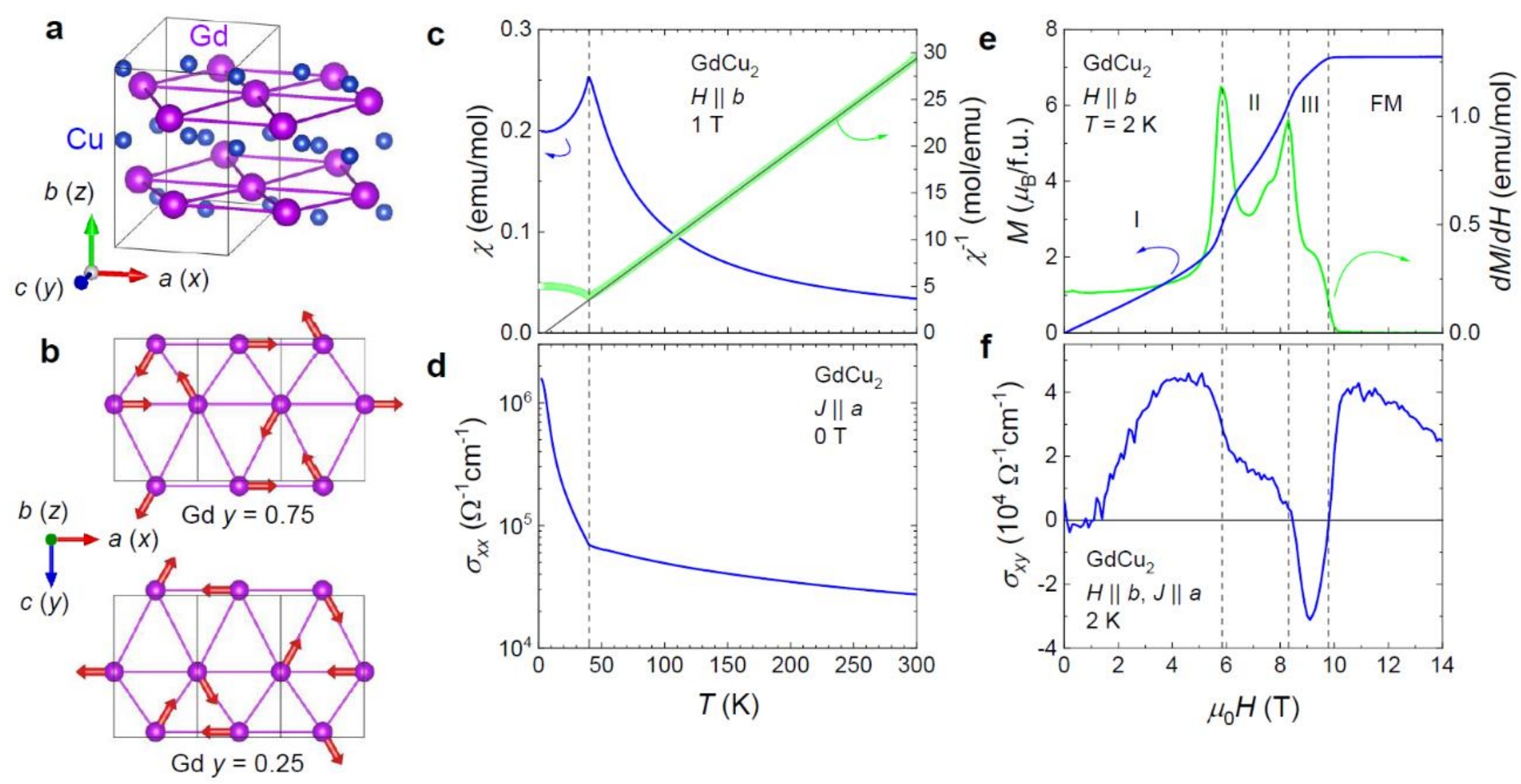

**Figure 1 | Crystal structure, and magnetic and transport properties of $GdCu_2$. a**, Schematic of crystal structure of $GdCu_2$. The black lines show an orthorhombic unit cell. **b**, Schematic of distorted triangular lattices of Gd atoms as viewed from the *b*-axis for the top layer ($y$ = 0.75) and the bottom layer ($y$ = 0.25). The magnetic moments with 120° cycloid propagation along the *a*-axis and anti-parallel stacking along the *b*-axis are drawn with red arrows. **c**, Temperature ($T$) dependence of magnetic susceptibility $\chi = M/H$ at 1 T (blue line). The inverse susceptibility $\chi^{-1}$ (green symbols) is fitted with the Curie-Weiss law above 150 K ($T_{CW}$ = 4.26 K, $\mu_{eff}$ = 8.97 $\mu_B$/Gd). **d**, Temperature dependence of longitudinal conductivity $\sigma_{xx}$ at zero field. **e**, Magnetic field ($H$) dependence of magnetization $M$ (blue line) and its field derivative $dM/dH$ (green line) at 2 K. **f**, Magnetic field dependence of Hall conductivity $\sigma_{xy}$ at 2 K.

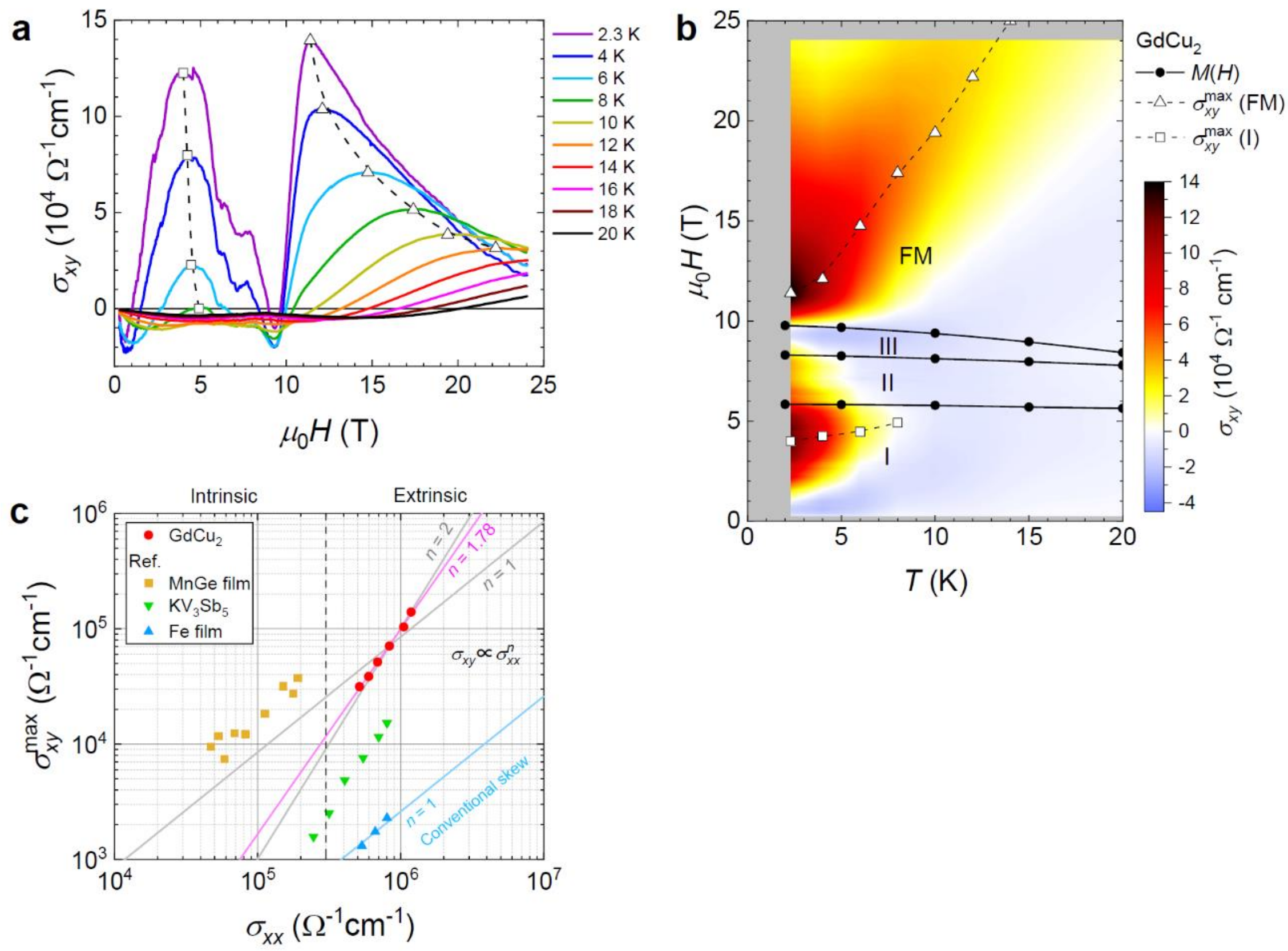

**Figure 2 | Temperature and field dependence of Hall conductivity of $GdCu_2$. a**, Magnetic field dependence of Hall conductivity at various temperatures. The square and triangle open symbols indicate the peak positions. **b**, Contour plot of Hall conductivity on the temperature-field phase diagram. The phase boundaries shown by the closed circle symbols are determined by magnetization measurements (Fig. 1e and Supplementary Fig. S2). The square and triangle open symbols indicate the peak positions of $\sigma_{xy}(H)$ in phase I and the FM region, respectively. **c**, Scaling plot of the maximum Hall conductivity $\sigma_{xy}{}^{max}$ in the FM region versus the longitudinal conductivity $\sigma_{xx}$ obtained at low temperatures from 2 K to 12 K. The pink line shows the fitting of the scaling law $\sigma_{xy} \propto {\sigma_{xx}}^n$ ($n$ = 1.78). For comparison, the conventional skew scattering of Fe films and the giant anomalous Hall conductivity of $KV_3Sb_5$ [14] and MnGe films [15] are also plotted.

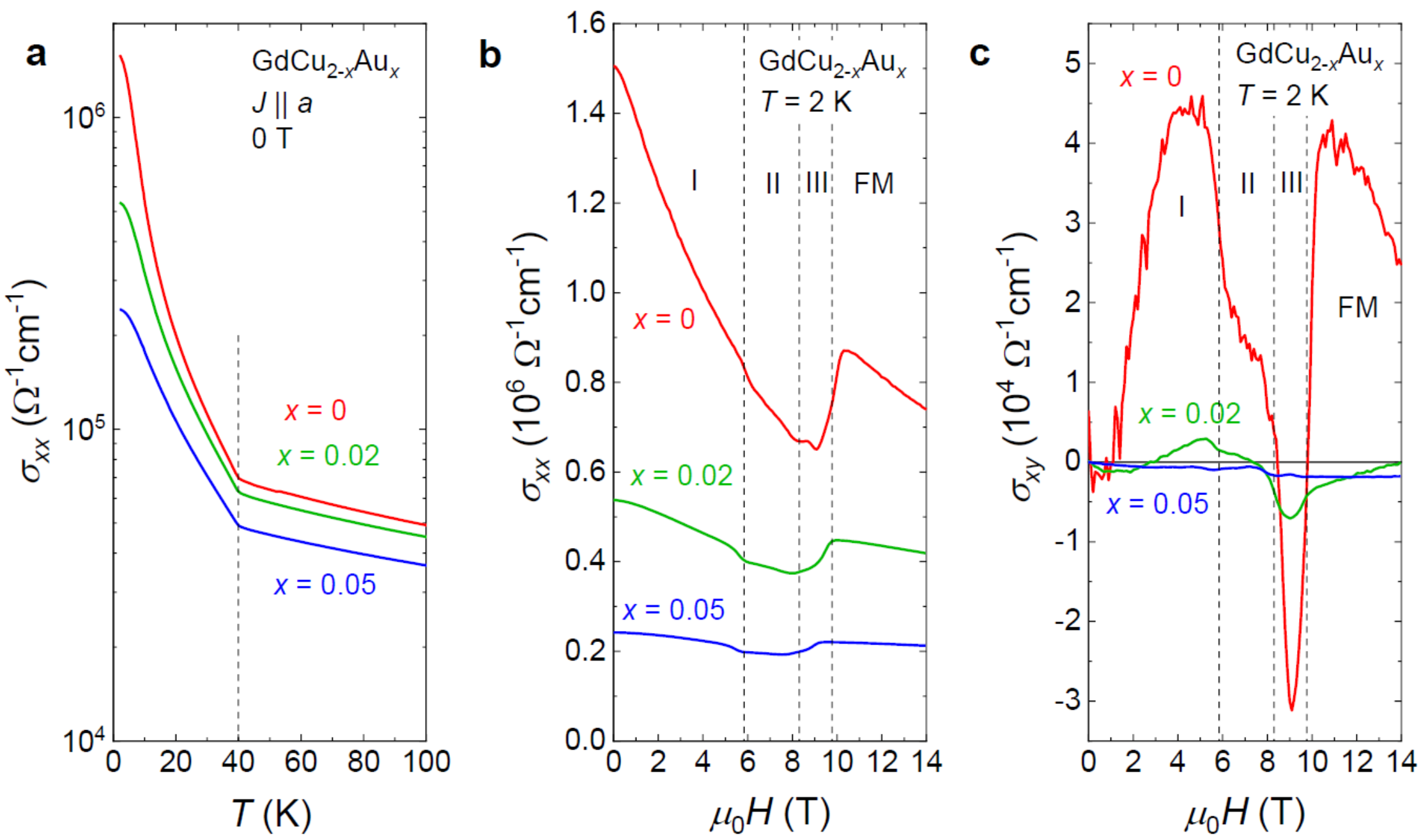

**Figure 3 | Au-doping effect on the conductivity in $GdCu_2$. a**, Temperature dependence of longitudinal conductivity of $GdCu_{2-x}Au_x$ ($x$ = 0, 0.02, 0.05) at zero field. **b**, **c**, Magnetic field dependence of (**b**) longitudinal and (**c**) Hall conductivity of $GdCu_{2-x}Au_x$ ($x$ = 0, 0.02, 0.05) at 2 K.

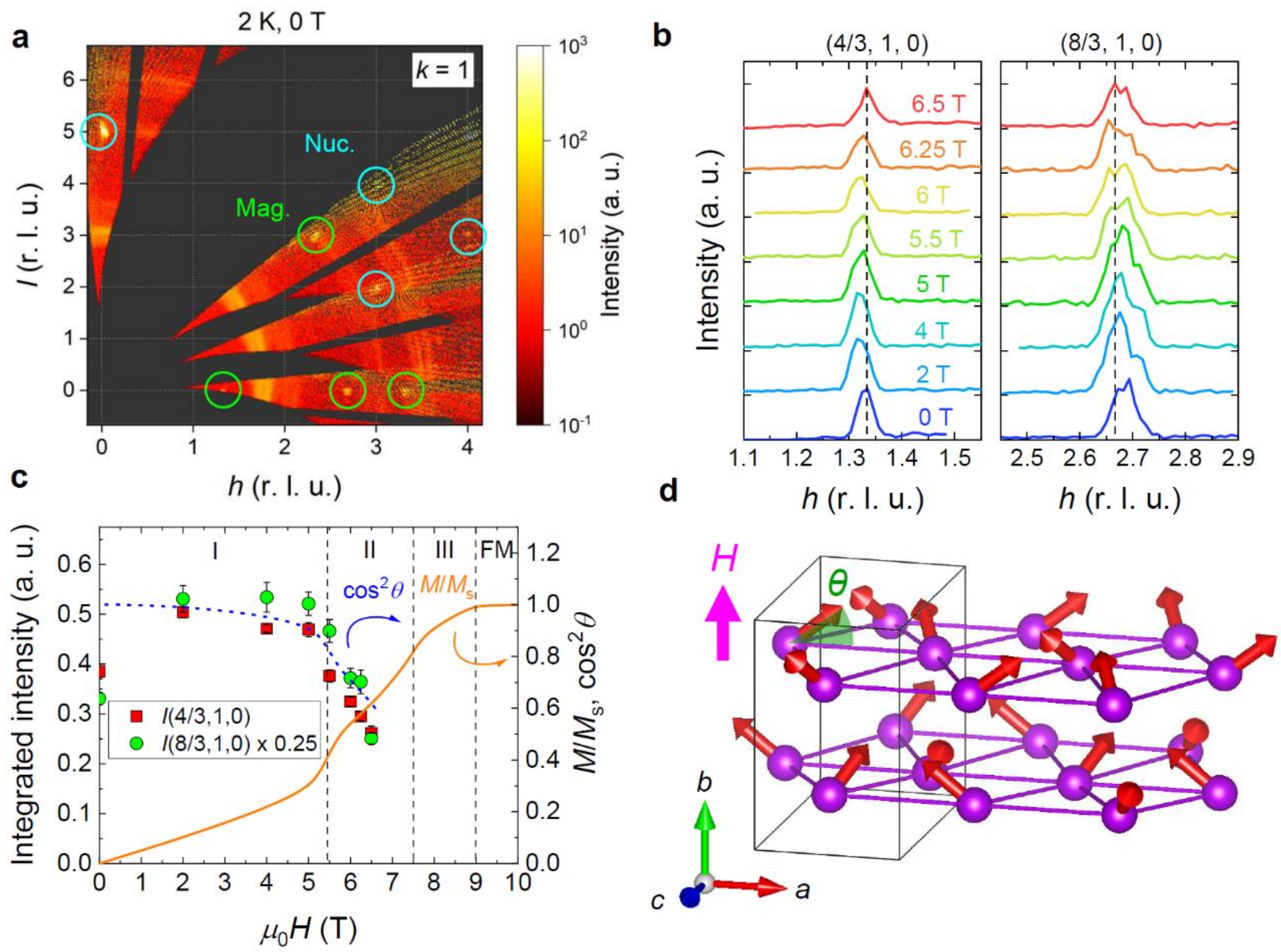

**Figure 4 | Magnetic structure of $GdCu_2$. a,** Contour map of neutron scattering intensity on the reciprocal lattice (*h*, 1, *l*) plane at 2 K and 0 T. The nuclear and magnetic peaks are indicated by light-blue and green circles, respectively. **b,** Neutron scattering intensity profile along the (*h*, 1, 0) line around the center of $h = 4/3$ and $h = 8/3$ at several magnetic fields from 0 to 6.5 T. For clarity, each data was shifted vertically by a constant interval. **c**, Magnetic field dependence of the integrated intensity of the (4/3, 1, 0) magnetic peak (red squares) and (8/3, 1, 0) magnetic peak (green circles). Magnetization normalized by the saturation value at 14 T ($M/M_s$) and the calculated $\cos^2\theta$ are also plotted with solid orange line and broken blue line, respectively. Here, $\theta$ [= $\arcsin(M/M_s)$] is defined as the angle of the Gd moments tilted from the *ac*-plane to the *b*-axis. **d**, Schematic of the magnetic structure of Gd local moments under magnetic fields.

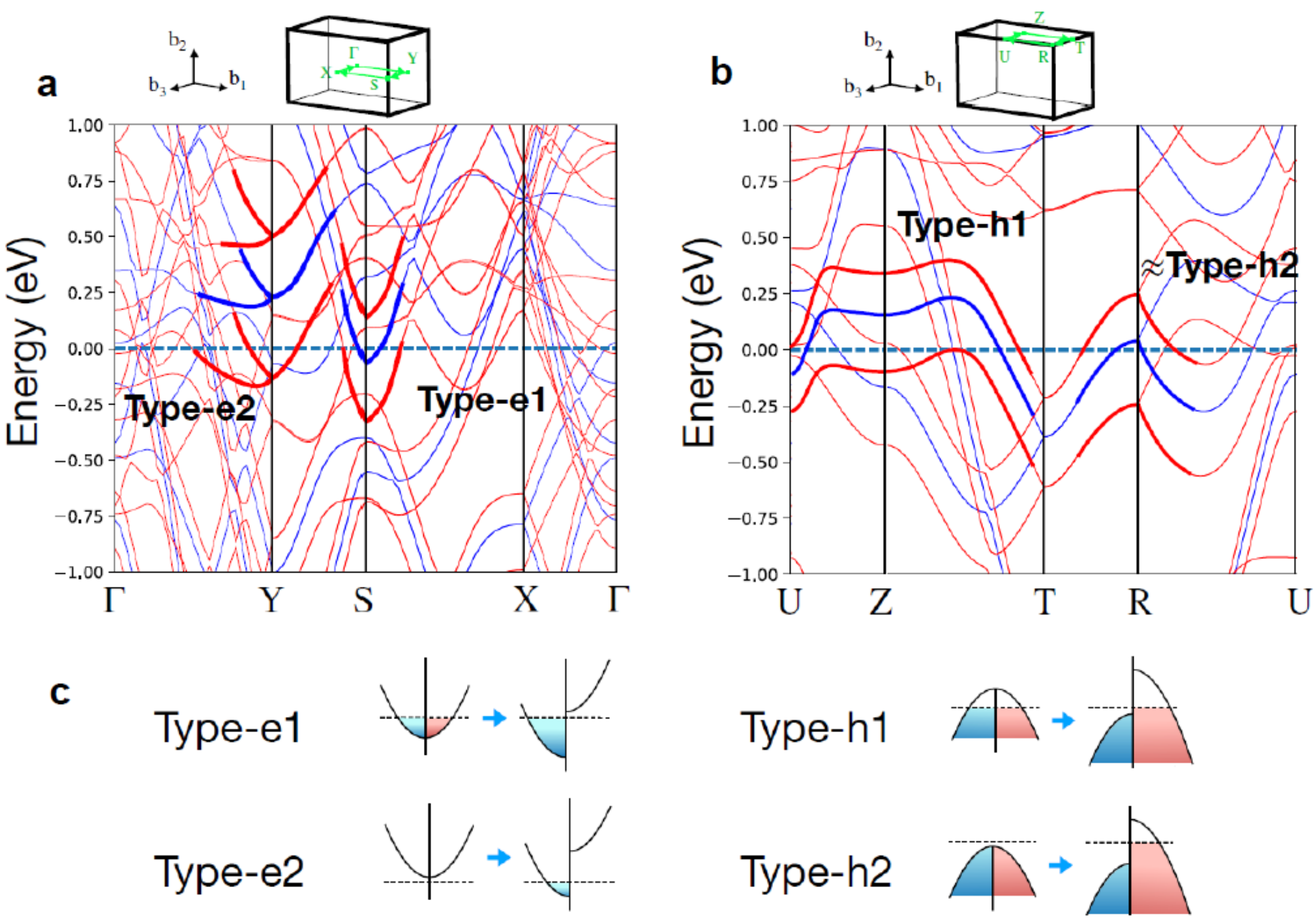

**Figure 5 | First-principles calculations for $GdCu_2$. a**, **b**, Band structure along the high-symmetry line for (**a**) $k_z = 0$ and (**b**) $k_z = 0.5$ in PM state (blue line) and FM state (red line). We highlight bands that substantially change the Fermi surface and potentially cause the anomalous behavior of the Hall conductivity. $E_F$ is taken at 0 eV. The corresponding Fermi surface is presented in Supplementary Fig. S9. **c**, We categorize the bands by their dispersion and their relative level compared to the $E_F$ with/without spin-splitting. In the cases of these four kinds of bands, carrier density has strong magnetization dependence.

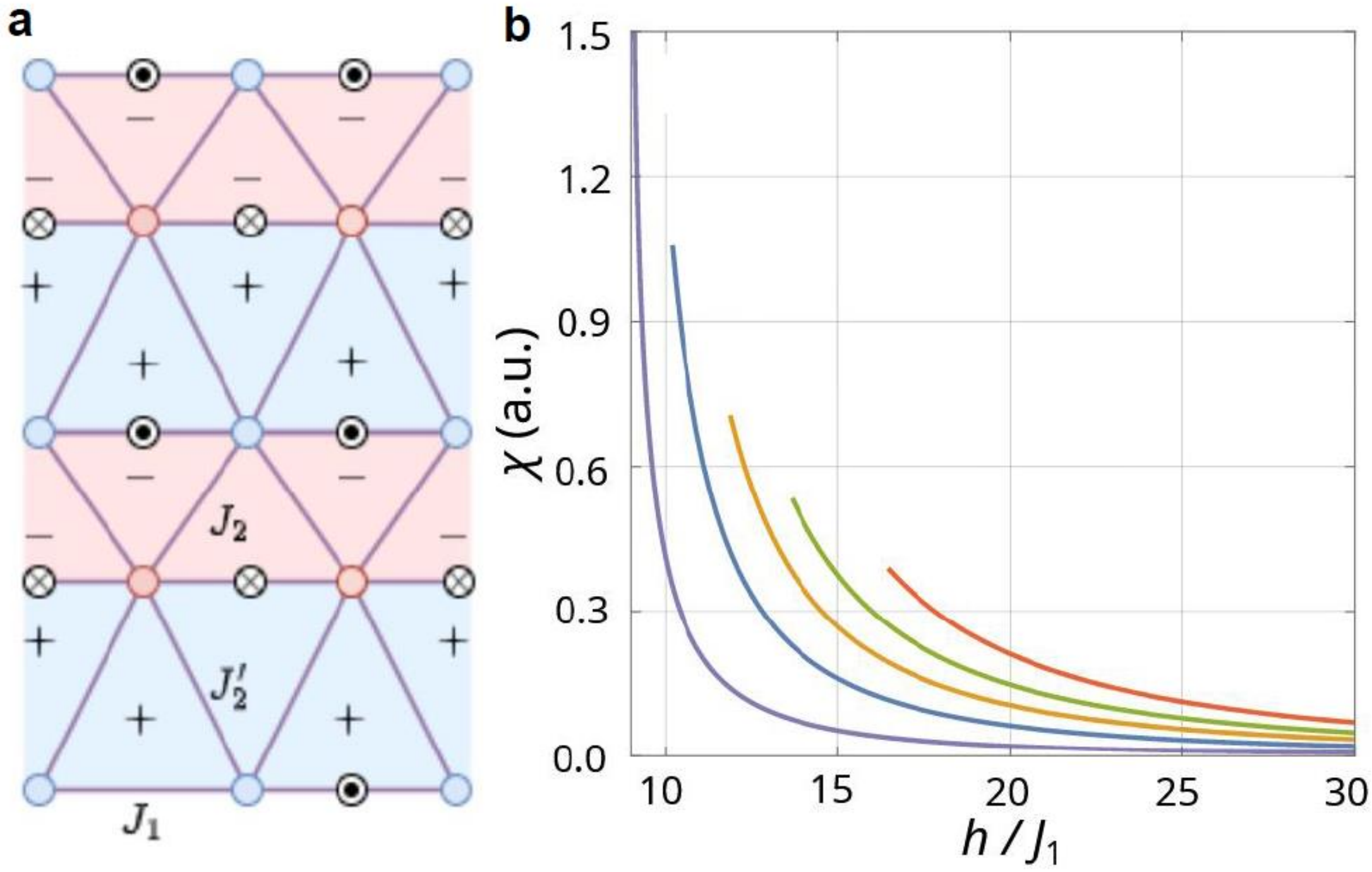

**Figure 6 | Effective model and spin chirality in the FM phase. a**, Schematic of the spin model. The circles with dots and crosses represent the direction of Dzyaloshinskii-Moriya vector and the plus and minus signs represent the signs of spin chirality when $D < 0$. **b**, Magnetic field ($h$) dependence of scalar spin chirality $\chi_1$ for $J_1 = J_2 = J_2' = -1$ and $D = -0.1$. Each line is for $T/|J_1| = 0.1$ (purple), 0.3 (blue), 0.5 (yellow), 0.7 (green), and 1.0 (red). The plots are for when the thermal average of transverse spin components is $\langle S_{ix}^2 + S_{iy}^2 \rangle \leq 0.1$.